\documentclass[twocolumn,showpacs]{revtex4}

\usepackage{graphicx} 
\usepackage{multirow,booktabs}      % advanced tables handling
\usepackage{amssymb,amsmath}        % advanced math extension
\usepackage{textcomp}               % textsymbols extension
\usepackage{indentfirst}            % name says all
\usepackage{hyperref}               % hypertext in pdf's
\usepackage{wasysym}
\usepackage{listings}

\begin{document}

  \title{Reducing the parameter space for Unparticle-inspired models using white dwarf masses}
  \author{Rodrigo Alvares de Souza}
    \email[Email: ]{rodrigo.souza@usp.br}
  \author{J.E. Horvath}
    \email[Email: ]{foton@astro.iag.usp.br}
  \affiliation{Instituto de Astronomia, Geof\'isica e Ci\^encias Atmosf\'ericas da USP
  			\\Rua do Mat\~ao 1226, Cidade Universit\'aria, 05508-090, S\~ao Paulo, Brazil}

	\date{\today}

  \begin{abstract}
	Based on astrophysical constraints derived from Chandrasekhar's mass limit for white-dwarfs,
	we study the effects of the model on the parameters of unparticle-inspired gravity, on scales
 $\Lambda_U > 1 \; TeV$ and $d_U \approx 1$.
		
\vskip 0.5cm

\end{abstract}


\maketitle

\section{Introduction}

Many proposals for explaining the apparent shortcomings of the Standard Model have been advanced.
The Unparticle Model proposed by Georgi \cite{bib:georgi_2007} aimed to include in the
Standard Model massive but scale-invariant particles, sharing the same physics of the scale-dependent counterparts.
These objects, called `unparticles', could play an important role in low-energy physics \cite{bib:PhysRevLett.100.031803},
since the model implies that unparticles can be exchanged between massive particles, leading to a new force
called `ungravity'. This ``fifth'' force would add a perturbation term to the newtonian gravitational potential, although the exact
potential can not be obtained because the distance at which the perturbed potential matches the
newtonian expression needs to be known. In order to bypass this limitation, the perturbed potential has been assumed to be of the
form \cite{bib:PhysRevLett.100.031803}

\begin{equation}
	V(r) = -\frac{GM}{2r}\left[ 1 + \left( \frac{R_G}{r} \right)^{2d_U-2} \right],
	\label{eq:newtonian_perturbed_potential}
\end{equation}

where $d_U$ (the scaling dimension of the unparticles operator $O_U$)
is $ \approx 1$, as a reasonable approximation, and $R_G$ is the characteristic
length scale of ungravity, given by

\begin{eqnarray}
	R_G & = & \frac{1}{\pi \Lambda_U} \left( \frac{M_{Pl}}{M_*} \right)^{1/(d_U - 1)} \times \nonumber\\
	& \times & \left[ \frac{2(2-\alpha)}{\pi} \frac{\Gamma(d_U + 1/2)\Gamma(d_U - 1/2)}{\Gamma(2 d_U)} \right]^{1/(d_U - 1)},
	\label{eq:R_G_definition}
\end{eqnarray}

where $\Lambda_U$ is the energy scale of the unparticle interactions, $M_{Pl} = 2.4 \times 10^{18} GeV$ is the Planck mass
and $\alpha$ is a constant dependent on the type of propagator considered.

The problem addressed in this work is to determine the bounds of the mass of the interaction
(un)particle $M_{*}$ with $d_U \simeq 1$. For this purpose, a suitable quasi-newtonian
gravitational system needs to be studied and compared with the pure newtonian results. A first study
of this regime by Bertolami, P\'aramos and Santos \cite{bib:bertolami_2009} has addressed the stellar equilibrium
problem, deriving a perturbed Lane-Emden equation further applied to the Sun. They explored the well-known
similarity of the full stellar structure to an $n = 3$ polytropic model, and derived limits
from the maximum allowed uncertainty in the central temperature $\Delta T_{c}/T_{c} = 0.06$.

In spite of the successful derivation of meaningful limits to the unparticle parameters, it is known that
the detailed structure of the Sun is actually quite complicated, and many physical factors have to
be considered beyond the simplest Chandrasekhar's polytropic model \cite{bib:chandrasekhar_1967}.
Therefore, it is worth considering another very well-known system to which the Chandrasekhar theory
gives an even better representation: the white dwarf sequence. We shall show below that an important
feature of these sequences (the maximum mass) is sensitive to the unparticle quantities and allows
to impose strong limits on them.

\section{Stellar Equilibrium and White-Dwarfs}
\label{sec:stellar_eq}
Since Chandrasekhar's polytropic model is widely known, we briefly recall how the unparticle theory
modifies it, as first shown by Bertolami, P\'aramos and Santos \cite{bib:bertolami_2009}.
The equations of stellar hydrostatic equilibrium and mass conservation can be reduced to a second order
differential equation, if a polytropic equation of state of the form $P=K \rho^{1+1/n}$ is assumed to hold.
If the density is written as $\rho = \rho_c \theta^n$, and the radius as $r = \beta \xi$, one can
easily obtain the original Lane-Emden equation

\begin{equation}
\frac{1}{\xi^2}\frac{d}{d\xi} \left( \xi^2 \frac{d \theta}{d\xi}\right) = -\theta^n,
\end{equation}

where $n$ is the polytropic index, $\xi$ is the dimensionless radius and $\beta$ is given by

\begin{equation}
	\beta= \left[\frac{(n+1)K}{4\pi G}\rho_c^{(1/n)-1}\right]^{\frac{1}{2}},
	\label{eq:lane_emden_parametro_beta}
\end{equation}

$K$ being the polytropic constant dependent on the specific value of $n$.
This equation is subject to the usual boundary conditions:
$\rho(r=0)=\rho_c$ and $dP/dr=0$ for $r=0$,
which translates to $\theta(\xi=0)=1$ and $\theta'(\xi=0)\equiv d\theta/d\xi = 0$. The detailed derivation
of the Lane-Emden equation can be consulted in the classical reference
\cite{bib:chandrasekhar_1967}. These results can be used
to derive the mass-radius relation, given by Eq.\eqref{eq:mass_radius_non_perturbed}.

\begin{equation}
	 m(r) = 4 \pi \left(\frac{r}{\xi_*} \right) ^\frac{3-n}{1-n}  \left(\frac{(n+1)K}{4 \pi G} \right)^\frac{n}{n-1} \xi^2 _* \left|\theta'(\xi_*)\right|
	\label{eq:mass_radius_non_perturbed}
\end{equation}

Considering that white-dwarfs are small stars composed by electron-degenerate matter, in which
the core material no longer undergoes fusion reactions, the Lane-Emden equation describes
very well their behavior as a result of the proximity of the electronic component to a polytropic
form. As is well-known, in the non-relativistic limit the white dwarf matter
can be represented by a $n=3/2$ polytrope and in the relativistic case the $n=3$ is quite accurate.

The same technique employed by Chandrasekhar can be used to obtain the perturbed Lane-Emden equation.
From the perturbed potential given by Eq.\eqref{eq:newtonian_perturbed_potential},
the perturbed gravitational acceleration can be obtained via $\vec{F} = -\vec{\nabla} V$
\cite{bib:carroll}, used to determine the hydrostatic equilibrium equation,
as seen in Eq.\eqref{eq:hydrostatic_equilibrium_perturbed}.

\begin{equation}
	 \frac{dP}{dr} = -\frac{G M \rho}{2r^2} \left[ 1 + (2 d_U - 1) \left( \frac{R_G}{r}\right)^{2d_U-2} \right]
	\label{eq:hydrostatic_equilibrium_perturbed}
\end{equation}

Then, from this hydrostatic equilibrium equation, Bertolami, P\'aramos and Santos
derived a perturbed Lane-Emden equation

\begin{eqnarray}
 \lefteqn{\frac{1}{\xi^2}\frac{d}{d\xi}\left(\xi^2 \frac{d\theta}{d\xi}\right)= } \nonumber\\
   && \frac{-\theta^n}{2}\left[ 1+(2d_U-1)\left(\frac{\xi_G}{\xi}\right)^{2d_U-2}\right] -\nonumber\\
   && {}-(2d_U-1)(d_U-1)\frac{1}{\xi}\frac{d\theta}{d\xi}\left(\frac{\xi_G}{\xi}\right)^{2d_U-2}
	\label{eq:lane_emden_perturbed}
\end{eqnarray}

in order to examine meaningful bounds on the parameters $R_G$ in
Eq.\eqref{eq:R_G_definition}, from which one can obtain the bounds of the mass of
the interaction (un)particle $M_{*}$ based on astrophysical constraints
(the parameter $\xi_G = R_G/\beta$ has been defined here to simplify the expressions).
We have corrected a slight misprint in the paper \cite{bib:bertolami_2009}
which would preclude the derivation of Eq.\eqref{eq:lane_emden_perturbed}, related
to the definition of the variable $\beta$ as given by
\eqref{eq:lane_emden_parametro_beta}.

Once these matters are settled, the perturbed mass-radius relation can be obtained
in the usual manner, being formally identical to the analogous mass-radius relation given by Chandrasekhar
(Eq.\eqref{eq:mass_radius_non_perturbed}), but with the perturbative effects present through
the solutions of the perturbed Lane-Emden equation and its derivative ($\theta(\xi_1)$ and $\theta'(\xi_1)$), 
respectively, both evaluated at the first zero as in Ref.\cite{bib:chandrasekhar_1967}. 
This modification sets the stage for
an analysis leading to novel bounds on the unparticle parameters.
Since we are interested on the maximum possible mass for a white dwarf, we set the parameters
related to the Lane-Emden model to the relativistic limit (polytropic index $n=3$),
and the white dwarf mass-radius relation becomes

\begin{equation}
	 M = 4 \pi  \left(\frac{K}{\pi G} \right)^\frac{3}{2} \xi^2 _* \left|\theta'(\xi_*)\right|
			\label{eq:mass_radius_non_perturbed_n_3}
\end{equation}

losing any dependence on the radius, as expected.

The key feature pointed out above, related to the explicit dependence of the maximum
mass through the zero of the Lane-Emden function and its slope, leads to quite a strong
dependence of the maximum mass with $d_U$ and $R_{G}$, which can be substantially different
from the ``canonical'' $M_{Ch}=1.457 (2/\mu_{e})^{2} M_{\odot}$
\cite{bib:chandrasekhar_1967, bib:landau_1965}.

\section{Observations and analysis}

Following the framework presented by Bertolami, P\'aramos and Santos , we obtained the
numerical solutions into Eq.\eqref{eq:lane_emden_perturbed} varying both parameters $\xi_G$ and $d_U$,
and used these solutions on Eq.\eqref{eq:mass_radius_non_perturbed_n_3}
to determine the masses associated to the combination of $d_U$ and $\xi_G$. It is important to note that
the combination $\xi_G$ and $d_U$ generated a broad range of masses, spanning from $M \approx 1.291 \; M_{\odot}$ to
$M \approx 1.874 \; M_{\odot}$. This in turn means that it is possible to constrain the parameters based on the maximum
mass for white dwarfs, because this mass must complain with the values obtained
by observational data. Following this reasoning, we assumed three reasonable values for the maximum mass

\begin{itemize}

\item the maximum observed masses in a large white dwarf sample, from the recent work by Kepler and collaborators
\cite{bib:2007MNRAS.375.1315K}, is $M=1.33 \; M_{\odot}$. Nevertheless, in order to effectively constrain the values of
$d_U$ near 1, we had to consider masses of at least $M=1.36 \; M_{\odot}$, otherwise $d_U$ would be far from 1
and the gravitational corrections in Eq.\eqref{eq:newtonian_perturbed_potential} would be too large;

\item the ``canonical'' limit, $M_{Ch}=1.457 \; M_{\odot}$ for carbon-type white dwarfs; in spite of
this widely accepted value is still beyond the actual observed maximum, its use seems very reasonable

\item an even larger value, above the former ``Chandrasekhar's'' limit, $M=1.60 \; M_{\odot}$,
arbitrarily chosen to represent an extreme limit allowed in nature.
If true, the number of objects between the actually observed
maximum $M=1.33 \; M_{\odot}$ and this proposed extreme value of $M=1.60 \; M_{\odot}$ must be
substantial in a large sample such as the one analyzed by Kepler et al. \cite{bib:2007MNRAS.375.1315K}, 
although none has been actually reported.

\end{itemize}

\begin{figure}[!ht]
  \centering
  \includegraphics[width=0.45\textwidth]{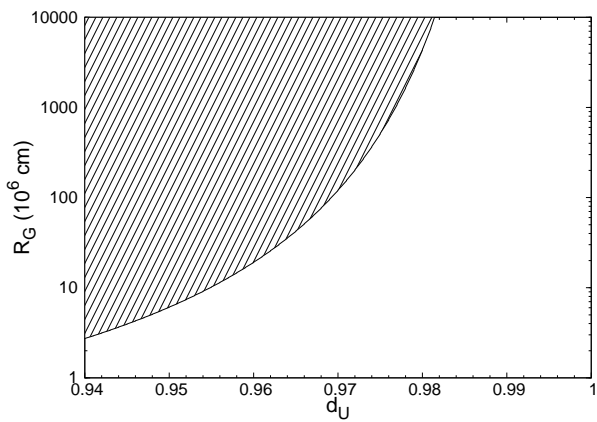}
  \caption{Contour plot ({\it locus}) of maximum mass ($M=1.6\;M_{\odot}$)
  in the $R_G$-$d_U$ plane, for an interaction
  range larger than the radius of the star. The dashed area
  gives masses bigger than $M=1.6\;M_{\odot}$, and should be considered as forbidden.}
  \label{fig:superior_range_2D_graph_RG}
\end{figure}

\begin{figure}[!ht]
  \centering
  \includegraphics[width=0.45\textwidth]{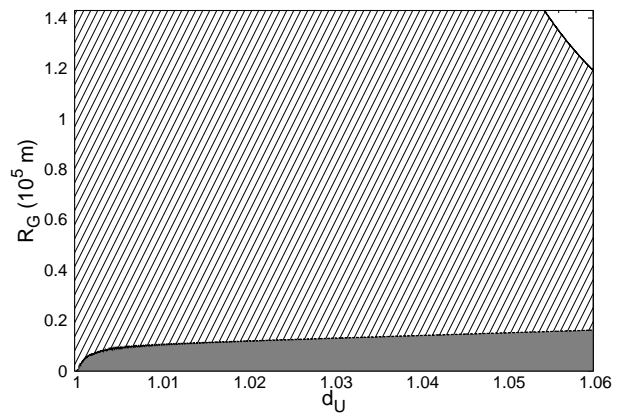}
  \caption{Contour plot ({\it locus}) of maximum masses $M=1.36\;M_{\odot}$ and $M=1.457\;M_{\odot}$ in
   the $R_G$-$d_U$, for an interaction
  range smaller than the radius of the star.
  The curve spanning from $1.055 < d_U < 1.06$ (upper right corner of the graphic) produces a maximum mass
  of $M=1.36\;M_{\odot}$ and the curve between $1 < d_U < 1.06$
  (lower side of the graphic) the mass of $M=1.457\;M_{\odot}$. The upper white area gives
  maximum masses lower than $M=1.36\;M_{\odot}$, the dashed area yields masses in the range $1.36\;M_{\odot}< M < 1.457\;M_{\odot}$,
  and the lower area produces masses bigger than $1.457\;M_{\odot}$. If one accepts the latter value as the maximum allowed, the 
  dark area is forbidden}
  \label{fig:inferior_range_2D_graph_RG}
\end{figure}

From the data above mentioned, we constructed the contour plots where
the combination of $d_U$ and $\xi_G$
generates the desired maximum mass. This plots are depicted in Fig.\ref{fig:superior_range_2D_graph_RG}
for ranges larger than the radius of the star ($\xi_G > 1$, $d_U \lesssim 1$) and in
Fig.\ref{fig:inferior_range_2D_graph_RG} for ranges smaller the radius of
the star ($\xi_G < 1$, $d_U \gtrsim 1$).
One should note that for $d_U \gtrsim 1$, only masses of $M=1.36\;M_{\odot}$
and $M=1.457\;M_{\odot}$ are obtained, while a mass of $M=1.6\;M_{\odot}$ can be imposed for $d_U \lesssim 1$ only.

\begin{figure}[!ht]
  \centering
  \includegraphics[width=0.45\textwidth]{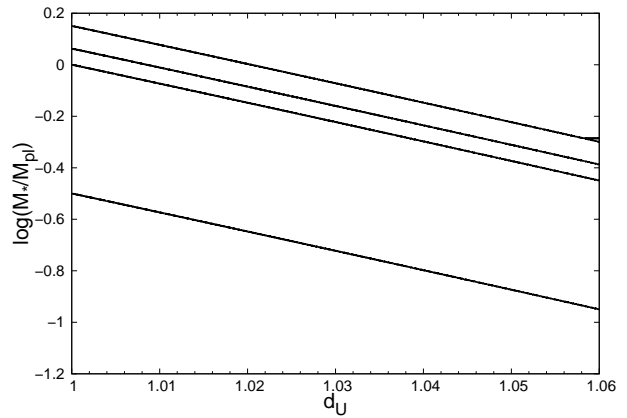}
  \caption{Lower bounds on $\log(M_{*}/M_{pl})$,
  for $\alpha=0$, $\alpha=2/3$, $\alpha=1$,
   $\alpha=1.9$, from top to bottom respectively. 
   The fixed values are $\Lambda_U=1 \; TeV$ and $M = 1.457 \; M_{\odot}$.}
  \label{fig:M_1.457_interno_Lambda_1TeV}
\end{figure}

\begin{figure}[!ht]
  \centering
  \includegraphics[width=0.45\textwidth]{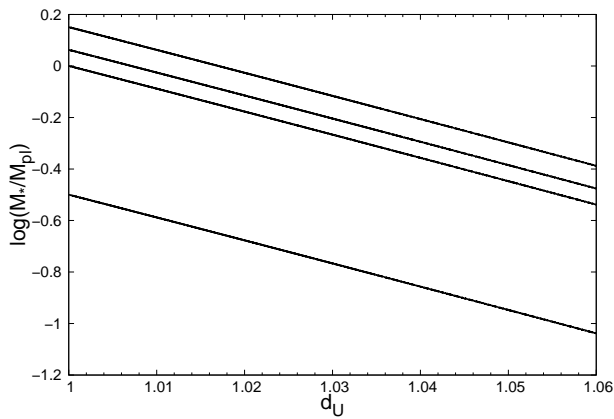}
  \caption{Lower bounds on $\log(M_{*}/M_{pl})$,
  for $\alpha=0$, $\alpha=2/3$, $\alpha=1$,
   $\alpha=1.9$ from top to bottom respectively. 
   The fixed values are $\Lambda_U=10^3 \; TeV$ and $M = 1.457 \; M_{\odot}$.}
  \label{fig:M_1.457_interno_Lambda_1E3TeV}
\end{figure}

The connection with the (un)particle mass follows by solving equation Eq.\eqref{eq:R_G_definition},
for the ratio $M_{*}/M_{Pl}$, namely

\begin{eqnarray}
	\frac{M_*}{M_{Pl}} & = & [\pi \Lambda_U R_G(d_U)]^{1-d_U} \times \nonumber\\
	 & \times &  \left[ \frac{2(2-\alpha)}{\pi} \frac{\Gamma(d_U + 1/2)\Gamma(d_U - 1/2)}{\Gamma(2 d_U)} \right]^{1/2},
	\label{eq:M_pl_M_star_definition}
\end{eqnarray}

which, by using $R_G(d_U)$ defined in the plots above, it is possible to infer the mass of the interaction (un)particle
$M_{*}$. We point out that this procedure differs significantly from the method used by Bertolami, P\'aramos and Santos
\cite{bib:bertolami_2009}, where the 6\% uncertainty leads to terms $R_{-}(d_U)$ and
$R_{+}(d_U)$, because we use exact values for the maximum masses. Nevertheless, it is clear that the ratio must be
interpreted as a lower bound to the ratio $M_{*}/M_{Pl}$, considering that lower maximum masses
(which would be in conflict with observations, as stated before) would result in even lower ratios.
Plotting the ratio as a function of $d_U$ and fixing the parameters $\alpha$ and $\Lambda_U$ in the same way
as Bertolami, P\'aramos and Santos, we obtain the ranges for the mass of the
interaction (un)particle depicted in Fig.\ref{fig:M_1.457_interno_Lambda_1TeV} and
Fig.\ref{fig:M_1.457_interno_Lambda_1E3TeV}
for the case of the ``canonical'' mass and two different choices of $\Lambda_{U}$.

Analyzing the data for $M = 1.36 \; M_{\odot}$, $d_U$s fall in the range $1.055 < d_U < 1.060$.
Therefore, this maximum mass yields a lower bound to $M_*$ in the range $(0.03 - 0.50)M_{pl}$.
For $M = 1.457 \; M_{\odot}$, $d_U$s lies in the range $1 < d_U < 1.06$, yielding a lower bound to $M_*$ in the range
$(0.1 - 1.6)M_{pl}$. The last case, with an assumed maximum mass $M = 1.6 \; M_{\odot}$
and the interaction range larger than the star radius, the only allowed values for
$d_U$s are in the range $0.94 < d_U < 0.982$, and lower bounds to $M_*$ in the range $(0.5 - 6.3)M_{pl}$.

\section{Conclusions}

We have shown in this work that quite strong limits to the unparticle
parameters can be obtained by using a simple form of the polytropic
theory of Chandrasekhar adding a perturbation to the Lane-Emden
equation, as first obtained by Bertolami, P\'aramos and Santos \cite{bib:bertolami_2009},
and applying it to the white dwarf sequences.

The key point elaborated here is that a change on the unparticle parameters would affect
the maximum mass allowed to white dwarfs, and thereby we explored this characteristic
in order to limit the values of such parameters.

The requirement that the maximum mass can not be too small (because it
would conflict with a few massive stars \cite{bib:2007MNRAS.375.1315K} or too big (because it
would lead to unobserved supermassive white dwarfs) limit the
values of $M_{*}$ to a confidence range of $0.1 M_{Pl} < M_{*} < 1.6 M_{Pl}$ from this analysis
alone for the case $d_U \gtrsim 1$. For the case $d_U \lesssim 1$, the mass-radius relation gives
only masses bigger than the canonical value. Considering that until today there is no observation of
white dwarfs with such high masses, this analysis may be interpreted to mean that
values of $d_U < 1$ are not allowed.

Following a different approach, based on a cosmological scenario, 
Bertolami and Santos \cite{bib:bertolami_santos_2009} considered the variation of the gravitational coupling
at the time of big bang nucleosynthesis, tensor exchange and the scaling dimension $d_U = 1.1$ and found $M_*$ to be
$> 0.05 M_{Pl}$, which is very close to the bounds found for $1 < d_U < 1.06$. 
Other works studying complementary bounds
\cite{bib:bertolami_2009, bib:PhysRevLett.99.141301, bib:1126-6708-2007-12-033, bib:Deshpande2008888, bib:JR2008561, bib:1475-7516-2009-03-019}
could be combined to address the viability of a general unparticle model, unless
one can manage to evade the bounds altogether. Even if so, a
general argument to constrain the admissible perturbations to the newtonian
potential can be made {\it via} the perturbed Lane-Emden equation, resorting
to the observed massive white dwarfs.

\begin{acknowledgments}

We wish to acknowledge the support of the CNPq and FAPESP Agencies for financial support.

\end{acknowledgments}

\end{document}